%% file: aa2872.tex
\documentclass[]{aa}
\usepackage{graphicx}
\usepackage{natbib}
\include{defs}
\begin{document}

\title{Very compact radio emission from high-mass protostars. \\
       II. Dust disks and ionized accretion flows.}

\author{F. F. S. van der Tak \and K. M. Menten}

\institute{Max-Planck-Institut f\"ur Radioastronomie, Auf dem H\"ugel
  69, 53121 Bonn, Germany \\ \email{vdtak@mpifr-bonn.mpg.de,kmenten@mpifr-bonn.mpg.de}}

\titlerunning{Very compact radio emission from high-mass protostars}
\authorrunning{Van der Tak \& Menten}

\date{Received 12 February 2005 / Accepted 22 March 2005}

\abstract{This paper reports 43~GHz imaging of the high-mass
  protostars W~33A, AFGL 2591 and NGC 7538 IRS9 at $\sim 0.04''$ and
  $\sim 0.6''$ resolution. In each case, weak ($\sim$mJy-level),
  compact (\O$\sim$100~AU) emission is detected, which has an
  elongated shape (axis ratio $\sim$3).
  However, for AFGL 2591 and NGC 7538 IRS9, the emission is
  single-peaked, while for the highest-luminosity source,
  W~33A, a `mini-cluster' of three sources is detected.
  The derived sizes, flux densities, and broad-band radio spectra
  of the sources
  support recent models where the initial expansion of \hii\ regions
  around very young O-type stars is prevented by stellar gravity. In
  these models, accretion flows onto high-mass stars originate in
  large-scale molecular envelopes and become ionized close to the
  star.  These models reproduce our observations of ionized gas as
  well as the structure of the molecular envelopes of these sources on
  10$^3$--10$^4$~AU scales derived previously from single-dish \smm\
  continuum and line mapping.
  For AFGL 2591, the 43~GHz flux density is also consistent with dust
  emission from a disk recently seen in near-infrared `speckle'
  images.  However, the alignment of the 43~GHz emission with the
  large-scale molecular outflow argues against an origin in a disk for
  AFGL 2591 and NGC 7538 IRS9. In contrast, the outflow from W~33A
  does not appear to be collimated.
  Together with previously presented case studies of W~3
  IRS5 and AFGL 2136, our results indicate that the
  formation of stars and stellar clusters with luminosities up to
  $\sim$10$^5$~\lsol\ proceeds through accretion and
  produces collimated outflows as in the solar-type case,
  with the `additional feature' that the accretion flow
  becomes ionized close to the star. Above
  $\sim$10$^5$~\lsol, clusters of \hii\ regions appear, and
  outflows are no longer collimated, possibly as the result
  of mergers of protostars or pre-stellar cores.


}

\maketitle

\section{Introduction}
\label{s:intro}

High-mass stars ($\gtsim$10~\msol) play a major role in the evolution
of their host galaxies, through strong ultraviolet radiation and
winds, and through supernova explosions. However, the origin of
high-mass stars is not well understood, due to the large distances
($\gtsim$kpc) and heavy extinctions ($A_V$$\gtsim$100) of the regions
they form in.  The optically visible main sequence life of OB-type
stars is preceded by an `embedded' phase which lasts $\approx$15\% of
their lifetime \citep{churchwell02:araa}.
Observations at mid-infrared through radio wavelengths have shown that
this embedded phase can be subdivided into several groups of objects: (1)
Infrared dark clouds, where internal density maxima and temperature
minima likely represent the initial conditions of high-mass star
formation. (2) High-mass protostellar objects, where the central star
is surrounded by a massive envelope with a centrally peaked
temperature distribution. (3) Hot molecular cores, with large masses
of warm and dense molecular gas, and large abundances of complex
organic molecules evaporated off dust grains. (4) Ultracompact H~II
regions, where large pockets of ionized gas have developed, but stay
confined to the stellar vicinity. (5) Compact, followed by classic
H~II regions, where the
ionized gas expands hydrodynamically and disrupts the parental
molecular cloud (for a review, see \citealt{gl99}).

Despite this progress, key questions remain unanswered, which
have been reviewed by \citet{churchwell02:boulder} and \citet{fvdt03}.
The paramount question is whether high-mass stars form via (disk)
accretion like low-mass stars, or that other mechanisms such as
coagulation of lower-mass stars or protostellar cores must
be invoked \citep{bbz98}. 
The ubiquity of massive \textit{bipolar} outflows, detected in 21 out
of a sample of 26 high-mass protostellar objects by \citet{beuther:outflow}
provides strong circumstantial evidence for disk accretion.
The second central question of high-mass star formation is the
relation with clustered star formation, i.e., the relation
between the mass distribution of a star-forming region and
its stellar density (e.g., \citealt{testi99}).
Addressing these questions requires \sas\ resolution, which at \smm\
wavelengths, with the Sub-Millimeter Array (SMA), is just coming
within reach \citep{beuther:orion}. However, \sas\ resolution can already
be achieved at centimeter wavelengths, where two important diagnostic
tools for high-mass star formation are continuum emission from dense
ionized gas and \hho\ maser emission.

Although the relation between the weak radio continuum emission
detected in some high-mass protostars to their overall luminosity is
presently not understood, it is clear that it marks a signpost for the
exact position of the object. For example, in the case of Orion-KL,
the weak radio source ``I'' almost certainly marks the position of the
exciting object in the region, as the excitation of the surrounding
SiO masers requires extreme temperatures and densities \citep{menten95}.
Apart from the signpost effect, the detection of radio emission, which
in the case of Orion-I is optically thick at frequencies up to at
least 345~GHz, can also put constraints on the nature of the embedded
protostar, in particular the accretion rate \citep{tan03}.

 Maser emission in the \hho\ $6_{16}$--$5_{23}$ line at 22.235~GHz
traces molecular gas with temperatures of $\sim$500~K and
$n$(\hh)$\sim$10$^9$~\ccm\ \citep{neufeld91} as occur for instance in
shocks. The high brightness of the maser emission permits high angular
resolution techniques and makes it a powerful tool for studying gas
motions on small scales.

We are exploring the diagnostic power of high-resolution cm-wave
observations to study the formation of high-mass stars.  In the source
AFGL 2136 \citep{mvdt04}, weak, compact radio continuum emission was
detected, as well as \hho\ maser emission which is redshifted from the
systemic velocity by $\approx$4~\kms. This emission may arise in
clouds of infalling gas that are heated in accretion shocks.
In another deeply embedded high-mass star-forming region, W3~IRS5,
\citet{claussen94} and \citet{tieftrunk97}
detected a cluster of seven compact radio continuum sources, some of which
are transient. Combining high-resolution millimeter-wave
and mid-infrared observations of these sources, \citet{keck05} conclude that
some of them are probably evaporating and/or shock-ionized clumps in the
ambient material, while others, having mid-infrared counterparts, most
likely contain young high-mass stars.

This paper presents new radio observations at high angular resolution of three
deeply embedded regions of high-mass star formation. The sources are a subset
of a sample of bright mid-infrared sources studied from the ground
(\citealt{willner82}; \citealt{mitchell:hot+cold}) and with ISO \citep{ewine98}.
Single-dish \smm\ continuum and molecular line mapping by \citet{fvdt00}
showed massive envelopes with a centrally concentrated temperature and density
structure.  For the present radio continuum study, we selected four sources
which cover a range of evolutionary states within the embedded phase, as
indicated by their \smm\ and infrared spectra \citep{sulph03}.

W~33A is a highly luminous object ($L$=1$\times$10$^5$~\lsol) in the
Galactic molecular ring. The kinematic distance based on CS and C$^{34}$S
lines is 4~kpc \citep{fvdt00}. Its massive, cold envelope
($\approx$1100~\msol\ within 74,000~AU radius) is a favourite target
for observations of mid-infrared absorption features due to icy grain
mantles \citep{gibb00}. The millimeter emission was resolved by
\citet{fvdt00} into a double source with separation 5$''$.
SMA observations will be presented by Shirley et al (in prep).
Two \hho\ maser spots are known, whose centroid lies 0\farcs75 from
the millimeter continuum peak at position angle --49$^\circ$ \citep{forster99}.
The maser velocities are redshifted by 0.2 -- 1.5~\kms\ relative to the
dense core.

AFGL 2591 is a relatively isolated luminous
($L$=2$\times$10$^4$~\lsol) protostar in the Cygnus~X region.
Distances in this region range from 0.5 to 2~kpc; we adopt $d$=1~kpc
and refer to \citet{fvdt99} for further discussion. The source is
known for its very high velocity CO mid-infrared absorption
\citep{mitchell:gl2591}. With an envelope mass of only $\approx$40~\msol\
within a 30,000~AU radius, it is probably the least embedded (most evolved)
source studied here.
A combination of compact radio continuum emission and \hho\
masers similar to AFGL 2136 has been observed by \citet{trinidad:gl2591}.
However, in AFGL 2591, the maser spots trace a helical structure
in position-velocity space, probably due to a precessing outflow.

NGC 7538 IRS9 is a luminous ($L$=4$\times$10$^4$~\lsol) source in the
NGC 7538 region, which optical spectrophotometry puts at
$d$=2.8~kpc \citep{crampton78}. This source combines strong
mid-infrared absorption features by icy grain mantles
with high velocity CO emission (\citealt{schutte96}; \citealt{mitchell:irs9}).
%\new{KMM: Need reference for high velocity CO!} 
The envelope mass of
$\approx$430~\msol\ inside a 66,000~AU radius makes it an intermediate
object between W~33A and AFGL 2591 in terms of envelope evolution.
\citet{kameya90} detected two \hho\ maser spots in NGC 7538 IRS9,
which have their centroid 0\farcs54 separated from the millimeter
continuum peak at position angle 40$^\circ$. Their
velocities are blueshifted by 4.7--17.1~\kms\ from the cloud velocity
and they probably arise in the outflow.

\begin{figure}[tbh]
\includegraphics[width=8cm,angle=0]{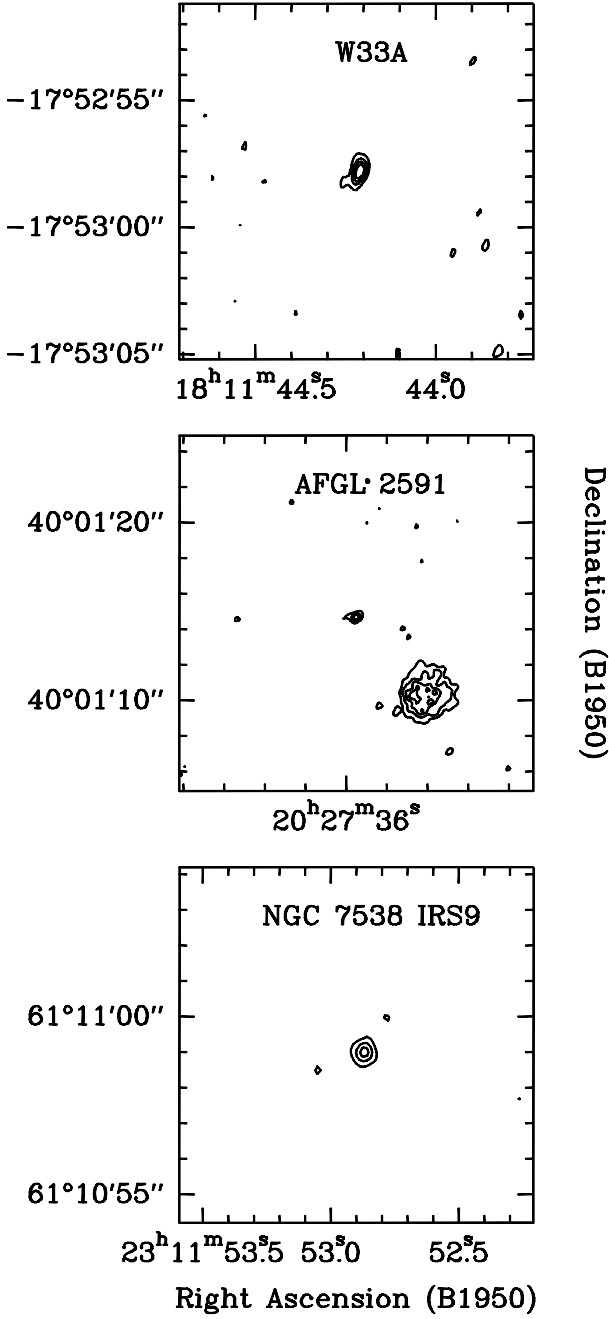}

\caption{VLA C-array 43~GHz images of our targets. Contours start at
  0.6 mJy/beam and increase by 0.6 mJy/beam for W~33A (top); for AFGL
  2591 (middle), contours start at 0.7 and increase by 0.7 mJy/beam;
  for NGC 7538 IRS9 (bottom), contours start at 0.5 and increase by
  1.0 mJy/beam.}

\label{f:cvla}
\end{figure}

\begin{figure}[tbh]
\includegraphics[width=8cm,angle=0]{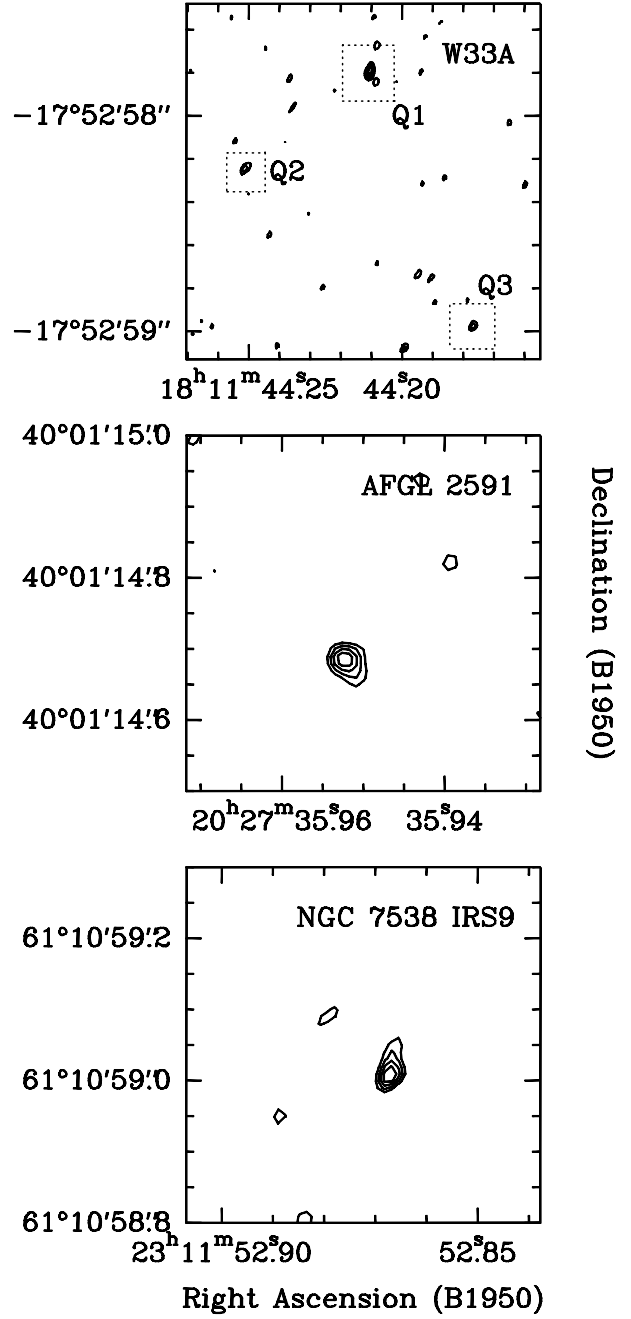}

\caption{VLA A-array 43~GHz images of W~33A (top), AFGL 2591 (middle)
  and NGC 7538 IRS9 (bottom). Contours start at 0.45 mJy/beam and
  increase by 0.15 mJy/beam. The boxes in the W~33A image are enlarged
  in the next figure.}

\label{f:avla}
\end{figure}

\begin{figure}[tbh]
\includegraphics[width=8cm,angle=0]{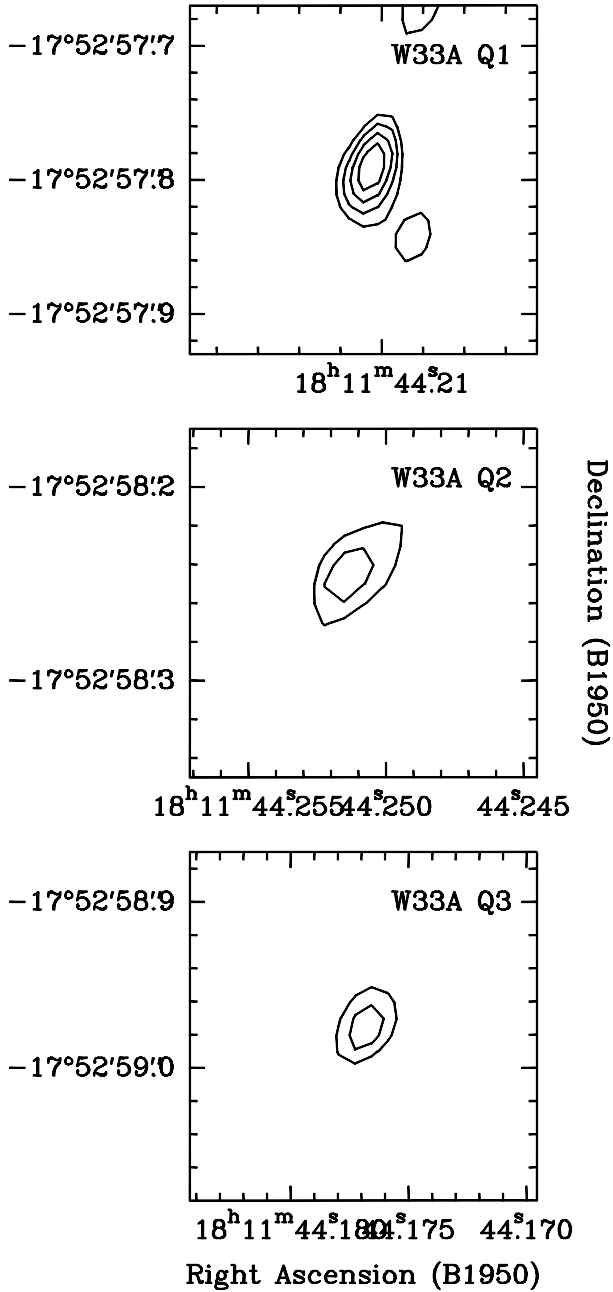}

\caption{Close-ups on the three 43~GHz sources in our VLA A-array
  image of W~33A. Contour levels are as in previous figure.}

\label{f:w33a}
\end{figure}

\section{Observations}
\label{s:obs}

Initial radio observations of W~33A, AFGL 2136, AFGL 2591, and NGC 7538
IRS9 were obtained on 2001 September 9, with the NRAO Very Large Array
(VLA)\footnote{The National Radio Astronomy Observatory (NRAO) is
  operated by Associated Universities, Inc., under a cooperative
  agreement with the U.S. National Science Foundation} in C--configuration.
Twenty-six antennas observed the sources sequentially 
in standard continuum mode:
two intermediate frequency (IF) pairs of width 50~MHz each
(43~MHz `effective' bandwidth) separated by 50~MHz.
The central frequency of 43.3~GHz is called Q--band by radio engineers.
Calibration of the antenna gains and phases was done by `fast
switching' to the sources 1817--254, 1829--106, 2013+370, and 2248+439
every 140 seconds.
%\new{KMM: What was the fast switching cycle time?}
The antenna pointing was checked every hour with a 1--minute
integration on the phase calibrators at a frequency of 8.4~GHz (`X--band').

Follow-up observations were made on 2002 March 23 and 25, with the
VLA in A--configuration. Twenty-three and twenty-four antennas
produced useful data on these respective days. The observing procedure
was the same as before, except that 1829--106 was used as phase
calibrator for both W~33A and AFGL 2136.

The data were edited, calibrated and imaged with NRAO's Astronomical
Image Processing System (AIPS). At the time of these observations,
corrections for elevation-dependent antenna gain and atmospheric
opacity were not applied automatically to VLA data, so we applied them
ourselves. In addition, small corrections to the positions of some
antennas were made.

The absolute flux density scale follows from observations of 1328+307
(3C286) which has a flux density of 1.46~Jy at this frequency
\citep{ott94}.  For the phase calibrators we obtain flux densities of
0.49$\pm$0.01~Jy for 1817-254, 0.63$\pm$0.01~Jy for 1829-106,
2.70$\pm$0.06~Jy for 2013+370 and 0.28$\pm$0.01~Jy for 2248+439. In
A--configuration, the primary flux calibrator, 3C286, is resolved, so
we did not use a point source model, but a detailed model (in the form
of Clean components) supplied by the VLA staff. We find flux densities
of 0.71$\pm$0.02~Jy for 1829-106, 3.61$\pm$0.12~Jy for 2013+370 and
0.29$\pm$0.01~Jy for 2248+439.

Figure~\ref{f:cvla} shows the C--array maps, obtained by a Fourier
transform of the $uv$ data with uniform weighting and deconvolution
with the Clean algorithm. These maps have rms noise levels of
0.23 mJy/beam. The synthesized beam sizes (FWHM) are 0.84$\times$0.41
arcsec at position angle (PA) $-$20$^\circ$ for W~33A,
0.53$\times$0.41 arcsec at PA $-$44$^\circ$ for AFGL 2591, and
0.55$\times$0.42 arcsec at PA $-$19$^\circ$ for NGC 7538 IRS9.
Figure~\ref{f:avla} shows the A--array maps, which have rms noise
levels of 0.18~mJy/beam. Restoring beam axes are 78$\times$36 mas at
PA $-$28$^\circ$ for W~33A, 43$\times$37 mas at PA $-$36$^\circ$ for
AFGL 2591, and 47$\times$36 mas at PA $-$42$^\circ$ for NGC 7538 IRS9.
Note that the regions shown in Fig.~\ref{f:avla} are about the size of
the compact emission sources from Fig.~\ref{f:cvla}.

\section{Results}
\label{s:res}

\begin{table*}
\caption{Gaussian fits to the observed radio emission. Numbers in brackets are
  uncertainties in units of the last decimal.}
\label{t:vla}

\begin{tabular}{llllllll}

\hline
\hline

Source        & $\alpha$ (B1950)  & $\delta$ (B1950)    & Peak $I_\nu$ & Total $S_\nu$ & Major axis & Minor axis & PA  \\
              & hh mm ss          & $^\circ$ $'$ $''$   & mJy/beam    & mJy          & mas           & mas        & deg \\
\hline

    & & & & \\

\multicolumn{8}{c}{\it C--array} \\

W 33A         & 18 11 44.2121(12) & $-$17 52 57.793(29) & 3.0(2)   & 4.3(5)  & 530(140) & 270(110) & 168(31) \\

AFGL 2591     & 20 27 35.9523(22) & $+$40 01 14.694(17) & 2.6(2)   & 3.3(5)  & 440(110) & $<$160 & 89(15) \\

NGC 7538 IRS9 & 23 11 52.8696(14) & $+$61 10 58.998(12) & 3.1(2)   & 2.9(3)  & 150(110) & $<$250 & 150(24)\\

    & & & & \\

\multicolumn{8}{c}{\it A--array} \\

W 33A Q1      & 18 11 44.2104(4)  & $-$17 52 57.804(9)  & 0.80(17) & 1.7(5)  &  87(57) & $<$67  &  41(20) \\
W 33A Q2      & 18 11 44.2510(3)  & $-$17 52 58.241(6)  & 0.79(18) & 0.62(27)&  $<$49  & $<$12  &   ...$^a$ \\
W 33A Q3      & 18 11 44.1769(3)  & $-$17 52 58.972(5)  & 0.76(18) & 0.53(25)&  $<$37  &  ...$^a$ & ...$^a$ \\

AFGL 2591     & 20 27 35.9520(3)  & $+$40 01 14.682(4)  & 0.95(16) & 1.9(5)  &  50(15) & 29(23) &  33(39) \\

NGC 7538 IRS9 & 23 11 52.8671(3)  & $+$61 10 59.017(4)  & 1.1(2)   & 1.1(3)  &  40(17) & ...$^a$ & 173(17) \\

\hline
\end{tabular}

\medskip

$^a$: No solution was found for this parameter

\end{table*}

\subsection{Positions and sizes}
\label{s:possiz}

%Morphology
Most images show single point sources. Only the C--array map of AFGL
2591 has an additional nearby extended \hii\ region. This source was
seen before by \citet{campbell84} at 4.9~GHz and by \citet{fvdt99} at
86~GHz, and appears partially resolved in our data. The flux density
in a box around this source is 67.6~mJy; the source is presumably
resolved out in A--array. We do not detect 
the somewhat extended
source VLA2, which \citet{trinidad:gl2591} only detect at
frequencies below 10~GHz. This source may be of non-thermal nature, 
or resolved out.

Table~\ref{t:vla} presents the results of Gaussian fits to the
observed compact radio emission, obtained with the task JMFIT in AIPS.
The A--array map of W~33A may show two weak sources in addition to the
source seen with the C--array.  We call the 43~GHz sources in W~33A
Q1, Q2 and Q3, in order of decreasing brightness.
Figure~\ref{f:w33a} shows close-ups of these sources.
If sources Q2 and Q3 are confirmed by more sensitive observations,
W~33A may resemble the W3~IRS5 region, where a cluster of radio
sources exists, some of which are internally, others externally
ionized \citep{keck05}.

%\bigskip

%Positions
The position of W~33A Q1 coincides with the 86~GHz
source MM1 from \citet{fvdt00}. The A--array image also shows an
extension of the emission toward the southwest, which was also seen
at 86~GHz and called MM2. Our data do not clearly show a second peak,
however. The \hho\ maser centroid \citep{forster99} lies 0\farcs44
from the 43~GHz peak, at position angle --78$^\circ$ (i.e., due West).
The positions of AFGL 2591 and NGC 7538 IRS9 are consistent with the
millimeter-wavelength (van der Tak et al.\ 1999, 2000)
%\citet{fvdt99} and \citet{fvdt00},
and mid-infrared positions.
% How far are the masers from the 43 GHz peak?
The positions found in A--array are consistent with those found in
C--array within their 2$\sigma$ errors.
%\bigskip

%Sizes
The fitted FWHM sizes of the sources are in columns 6 and 7, and the
orientation in column 8 of Table~\ref{t:vla}. The given sizes are
deconvolved values, even though most sources are only resolved in one
direction. In the few cases where no meaningful estimate of the minor axis
could be obtained, we use the value for the major axis as upper limit.
Using the distances from \S\ref{s:intro}, the A--array sizes
correspond to linear radii of 150~AU for W~33A, 20~AU for AFGL 2591,
and 55~AU for NGC 7538 IRS9.

The sizes found from the A--array data are not consistent with the
C--array estimates, but factors of 5--7 smaller. This discrepancy
suggests that the sources do not have a uniform brightness, but rather
a centrally peaked brightness distribution.
Each array picks up the `tip of the iceberg' at its own level of
brightness sensitivity.
%
%% The observational limits from the A--array data correspond to linear
%% sizes of 212~AU for W~33A, 40~AU for AFGL 2591, and 115~AU for NGC
%% 7538 IRS9.

The derived axis ratios (or limits) of $\ltsim$0.5 suggest a
nonspherical, flattened geometry. The position angles found from the
A--array images are consistent with those found from the C--array
images within their 2$\sigma$ errors.
Combining the results from both array configurations, the orientation
on the sky is PA$\approx$20$^\circ$ for W~33A Q1, PA$\approx$73$^\circ$
for AFGL 2591 and PA$\approx$165$^\circ$ for NGC 7538 IRS9.

\subsection{Flux densities and brightness temperatures}
\label{s:flux}

%Flux densities
Columns 4 and 5 of Table~\ref{t:vla} give the peak intensities and the
integrated flux densities of the sources.
The flux densities determined from the A--array data are $\approx$two
times lower than those derived from the C--array data. Since the phase
calibrators are 
%on average equally bright in both configurations, 
not systematically brighter in either configuration,
we rule out a systematic calibration offset between the two data sets.

Sources can be weaker in more extended configurations because they are
resolved, or because of increased phase noise on longer baselines.
This `atmospheric decorrelation' decreases the visibility amplitudes
by a factor of $e^{-\sigma_\phi^2 / 2}$, where $\sigma_\phi$ is the
phase noise (in radians). From the phase noise in the calibrator data,
we estimate a decorrelation by $\approx$5\% in C--array and by
$\approx$10\% in A--array. Since the discrepancy is much larger than
these factors, we conclude that decorrelation is unimportant here.
Combined with the difference in observed sizes (\S~\ref{s:possiz}),
this discrepancy in flux density is additional evidence for a
centrally peaked brightness distribution for these sources.

\bigskip

%Brightness temperature

A first clue to the nature of the emission is the conversion of the
measured flux densities and deconvolved sizes to brightness
temperatures, using the Rayleigh-Jeans law. The total flux densities
measured with the C--array correspond to $T_B$= 20 -- 50~K, while the
values measured in A--array indicate $T_B$=190 -- 850~K. Most of these
numbers are underestimates of the true brightness because the emission
is not fully resolved,
%Since the Rayleigh-Jeans limit is invalid for the C--array data 
so we adopt $T_B$ = 500 -- 1000~K as
representative range for these sources.
%\new{KMM: What do you mean with the R-J limit is ``invalid''?}

If due to ionized gas, which has an intrinsic temperature of 10,000~K,
the emission must be beam diluted, or optically thin, or both.
However, the cm-wave spectra of W~33A and AFGL 2591
(\S~\ref{s:spectra}) are too steep to be due to optically thin
free-free emission. Therefore, if the observed emission is from
ionized gas, the filling factor must be $\sim$0.05. Using the source
sizes derived in \S~\ref{s:possiz}, this filling factor implies
a linear radius  of
$\approx$30~AU for W~33A,
$\approx$5~AU for AFGL 2591 and
$\approx$10~AU for NGC 7538 IRS~9.

If the 43~GHz emission is due to dust, the maximum physical temperature is
$\approx$1500~K, where dust sublimates. Therefore the emission must be
close to optically thick and close to resolved. The mm-wave spectra
(\S~\ref{s:spectra}) indicate optically thick emission in W~33A and AFGL 2591,
but optically thin emission in NGC 7538 IRS9.
The filling factor must be at least 1/3 and is probably close to unity, especially if
the emitting dust is cooler than 1500~K. The corresponding physical
sizes assuming single condensations are
90 -- 150~AU for W~33A,
5 -- 20~AU for AFGL 2591, and
10 -- 30~AU for NGC 7538 IRS~9.

\subsection{Radio spectra}
\label{s:spectra}

\begin{figure}[tbh]
\includegraphics[width=7cm,angle=0]{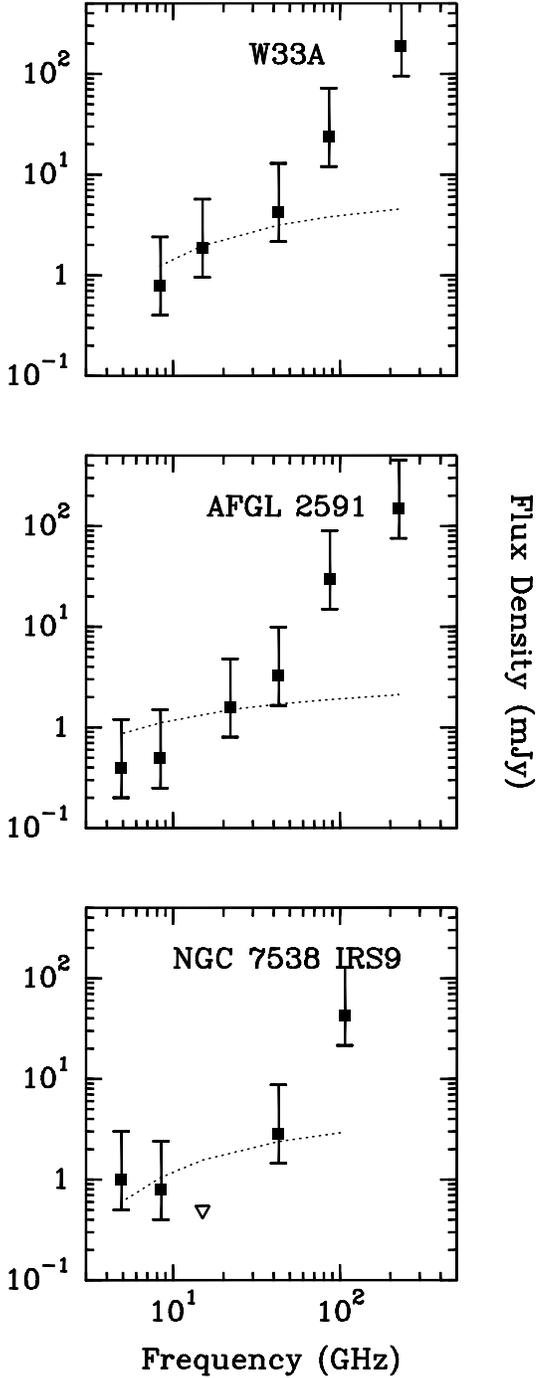}

\caption{Filled squares: Small-scale ($<$3$''$) radio (3--300~GHz)
  spectra of W~33A (top), AFGL 2591 (middle) and NGC 7538 IRS9
  (bottom). The `ionized accretion flow' model from \S~\ref{s:keto}
  (dotted lines) is seen to reproduce the emission up to $\sim$50~GHz
  to within a factor of two. Flux densities at higher frequencies
  $>$50~GHz are dominated by dust emission, which the model does not
  consider.}

\label{f:spec}
\end{figure}

Table~\ref{t:spec} contains radio flux densities for our three sources
from the literature. Only data between frequencies of 3--300~GHz and
in beams $\ltsim$3$''$ have been included. Figure~\ref{f:spec} plots
the data from Tables~\ref{t:vla} and~\ref{t:spec}. We use the total
flux densities measured in C--array, which are most comparable to the
literature data.

%spectral index
The broad-band spectrum of W~33A at frequencies up to 43~GHz can be
characterized by a spectral index $\gamma$, defined as
$$
S_\nu \propto \nu^\gamma
$$
with $\gamma$=1.2$\pm$0.4, while the spectrum steepens to
$\gamma$=2.3$\pm$0.2 at higher frequencies. Similarly, for AFGL 2591,
$\gamma$=0.9$\pm$0.4 at $\nu < 43$~GHz, while $\gamma$=2.5$\pm$0.8 at
$\nu > 43$~GHz. The spectrum of NGC 7538 IRS9 is even steeper,
$\gamma$=3.0$\pm$0.5, at $\nu > 43$~GHz, and inverted
($\gamma$=$-$0.5$\pm$0.4) at centimeter wavelengths.
The uncertainties in these spectral indices are dominated by the
differences in values derived at different frequencies, rather than by
the uncertainties of the individual flux densities (estimated as 10\%
for the VLA, 30\% for OVRO at 86~GHz and 50\% at 230~GHz).
The spectral indices may be affected by variability, since the
measurements were not simultaneous. In the case of W3~IRS5,
\citet{keck05} found radio variability by factors up to 3.

%\bigskip

The spectral indices of W~33A and AFGL 2591 indicate that the emission
at $\nu \ltsim 43$~GHz is dominated by free-free emission, while the
higher-frequency points are mostly due to dust emission. Similar
conclusions hold for AFGL 2136 and Orion-I, which indeed have similar
radio spectra \citep{mvdt04}.

%% \cut{In contrast, NGC 7538 IRS9 appears to have a non-thermal cm-wave spectrum
%% consistent with synchrotron emission from a jet, as seen before in
%% W3(\hho ) by \citet{reid95}. The similar flux densities of NGC 7538
%% IRS9 and W3(\hho ), which lie at comparable distances, lends further
%% support to a common emission mechanism. The mm-wave spectrum of NGC
%% 7538 IRS9 also differs from that of the other sources: the dust
%% emission in NGC 7538 IRS9 must be optically thin.}

The spectrum of NGC 7538 IRS9 differs from that of the other sources
in two respects. First, the mm-wave spectrum is significantly steeper
than $\gamma$=2, indicating that the dust emission in NGC 7538 IRS9 is
optically thin. 
%\new{KMM: Is that really true?}
Second, NGC 7538 IRS9 has a `flattish' spectrum between 4.9 and
8.5~GHz, whose spectral index is consistent with 0, but poorly
determined, due to the limited signal-to-noise ratio in particular of
the 8.5~GHz emission \citep{sandell05}.  The latter authors interpret
this as free-free emission, which, given the spectral behaviour would
have to be optically thin, which is unlikely for such a compact
source. The sizes they determine for the radio source, $2''\times1''$
at 8.46 GHz and $0.7''\times0.1''$ at 4.86 GHz, also are very
uncertain, and should probably at best be considered as upper limits
(consistent with our 43 GHz data).  We note however, that within the
uncertainties, NGC 7538 IRS9's radio spectrum could be due to
non-thermal emission, consistent with synchrotron emission from a jet,
as seen before in the `Turner-Welch' object W3(\hho ) by
\citet{reid95}. An origin of the radio emission in a jet is consistent
with the source's morphology, which is unresolved in one dimension.
Its orientation, almost north-south, is similar to that of the
high-velocity CO outflow \citep{sandell05}, although we note that
outflows around NGC 7538 IRS9 observed in other tracers (\hh\ and
HCO$^+$) present a complex picture.
%% \cut{The similar flux densities of NGC 7538
%% IRS9 and W3(\hho ), which lie at comparable distances, lends further
%% support to a common emission mechanism.}

The centimeter-wave spectral indices of W~33A and AFGL 2591 are
intermediate between the values expected for optically thin
($\gamma$=--0.1) and optically thick ($\gamma$=2) \hii\ regions with a
uniform temperature and density. An intermediate optical depth would
lead to a `bent' spectrum which is not observed. However, the observed
values of $\gamma \approx 1$ are expected in the case of \hii\ regions
with density gradients.  The most plausible gradients are the cases of
an ionized wind and an infalling envelope, which lead to $n\propto
r^{-2}$ and $n\propto r^{-1.5}$ density structure, respectively. The
expected spectral indices of $\gamma$=0.6 and $\gamma$=0.1 are
somewhat below the observed values.

In the case of a power-law distribution of the electron density with
radius, $n_e \propto r^{-q}$, a spectral index of $\gamma =
(2q-3.1)/(q-0.5)$ is expected \citep{olnon75}. Therefore the observed
value $\gamma \approx 1$ would indicate $q=2.6$. However, it is not
clear which process would create and sustain such a steep density
distribution. The sound speed of $\approx$10~\kms\ implies a crossing
time of $\approx$50~yr for our sources, so the process would have to
act continuously. More plausible models will be discussed in
\S~\ref{s:keto}.

\begin{table}
\caption{Overview of literature data.}
\label{t:spec}

\begin{tabular}{lrrc}

\hline
\hline

Source        & $\nu$ & $S_\nu$ & Reference \\
              & (GHz) & (mJy)   &           \\
\hline

W 33A         & 8.4   & 0.8     &  4 \\
              &  15   & 1.9     &  4 \\
              &  86   &  24     &  5 \\
              & 233   & 190     &  5 \\

AFGL 2591     & 4.9   & 0.4     &  1 \\
              & 8.4   & 0.5     &  2 \\
              &  22   & 1.6     &  6 \\
              &  87   &  30     &  3 \\
              & 226   & 151     &  3 \\

NGC 7538 IRS9 & 4.9  & 1.0    & 7 \\
              & 8.5  & 0.8    & 7 \\
              & 15   & $<$0.5 & 4 \\
              & 107  & 43     & 5 \\

\hline
\end{tabular}

\medskip

References:
(1) \citealt{campbell84}
(2) \citealt{tofani95}
(3) \citealt{fvdt99}
(4) \citealt{rengarajan96}
(5) \citealt{fvdt00}
(6) \citealt{trinidad:gl2591}
(7) \citealt{sandell05}

\end{table}

\section{Discussion}
\label{s:disc}

\subsection{Stellar ionization rates}
\label{s:ion}

The flux densities $S_\nu$ in Table~\ref{t:vla} can be combined
with the distances $d$ from \S~\ref{s:intro} to estimate the flux of Lyman
continuum photons $N_L$ emitted by the central star (e.g., \citealt{kruegel}).
In the optically thick limit, the `Str\"omgren' radius $r_S$ of the
\hii\ region follows from blackbody emission at an assumed electron
temperature of $T_e$=10$^4$~K:

$$
S_\nu = B_\nu (T_e) \frac{r_S^2}{d^2} \frac{\pi}{\ln 2}
$$

The derived radii of 33.5~AU for W~33A, 7.3~AU for AFGL 2591, and
19.3~AU for NGC 7538 IRS9 are consistent with the
observational limits (\S~\ref{s:possiz}).
To have a free-free optical depth of unity, the emission measure must
be at least 8.4$\times$10$^9$ cm$^{-6}$\,pc.
These radii and emission measures together imply electron densities of
$n_e$=5$\times$10$^6$ to 1$\times$10$^7$~\scm.
The ionizing photon flux then equals the recombination rate:

$$
N_L = \frac{4\pi}{3} r_S^3 n_e^2 \alpha_B
$$

where $\alpha_B$ is the `case~B' recombination coefficient of
2.59$\times$10$^{-13}$ cm$^3$s$^{-1}$ \citep{osterbrock}. The values
of $N_L$ are 3.5$\times$10$^{45}$ s$^{-1}$ for W~33A,
1.7$\times$10$^{44}$ s$^{-1}$ for AFGL 2591, and 1.2$\times$10$^{45}$ s$^{-1}$
for NGC 7538 IRS9.
Values estimated using the optically thin limit,

$$ N_L = 7.54 \times 10^{46} \frac{S_\nu}{{\rm Jy}}
      \left( \frac{d}{{\rm kpc}} \right)^2
      \left( \frac{T_e}{10^4 {\rm K}} \right)^{-0.45}
      \left( \frac{\nu}{{\rm GHz}} \right)^{-0.1}
$$
are within a factor of 2 from the thick estimates.

\subsection{Dust accretion rates}
\label{s:stevol}

Ionizing photon fluxes can also be estimated from the infrared
luminosities, assuming that one single star dominates the Lyman
continuum flux. We use the Geneva stellar structure models
\citep{maeder89} to convert luminosities into stellar masses and
effective temperatures, and the stellar atmosphere models by
\citet{sdk97} to estimate Lyman continuum fluxes.
The results are
$M$=28~\msol, \teff=40,000~K and $N_L$=8.5$\times$10$^{48}$ s$^{-1}$ for W~33A,
$M$=16~\msol, \teff=33,000~K and $N_L$=1.0$\times$10$^{48}$ s$^{-1}$ for AFGL 2591, and
$M$=20~\msol, \teff=35,000~K and $N_L$=2.2$\times$10$^{48}$ s$^{-1}$ for NGC 7538 IRS9.

These ionizing photon fluxes are factors of 2000--6000 higher than the estimates
from the radio continuum emission (\S~\ref{s:ion}). Dust inside the \hii\ region only
accounts for factors of 2 -- 3 absorption.
One possibility is confinement by the ram pressure of accreting dust
(cf.\ \citealt{walmsley95}).
The required accretion rate follows from equating the ram pressure to
the thermal pressure of the \hii\ region:
$$
\dot{M}v_{\rm ff} = 4\pi r_S^2 n_ek_BT_e
$$
where the free-fall velocity is given by
$$
v_{\rm ff} = \sqrt{GM_* / r_S}
$$
which equals 42~\kms\ for $M_*$=20~\msol. With $r_S$=20~AU and
$n_e$=10$^7$~\ccm, the estimated accretion rate is 6$\times$10$^{-8}$ \msol~yr$^{-1}$.
This value is a lower limit because radiation pressure may slow down
the dust from the free-fall speed, and because a stellar wind may add
to the internal pressure.

\subsection{Ionized accretion flows}
\label{s:keto}

Recent work by \citet{keto02} shows that stellar gravity prevents the
hydrodynamical expansion of \hii\ regions inside a `gravitational
radius'

$$
r_g = GM / 2 c_s^2
$$

where $c_s$ is the isothermal sound speed of ionized gas of
$\approx$10~\kms. This radius lies at 124~AU for W~33A, 71~AU for AFGL
2591, and 89~AU for NGC 7538 IRS9. These values are comparable to the
observational limits (\S~\ref{s:possiz}).

In Keto's model, gas accreting onto the star is in molecular form at
large radii, and becomes ionized at $r=r_g$. The density structure of
the molecular and ionized regions is given by the $n\propto r^{-1.5}$
free-fall profile. The molecular envelopes of our sources have similar
or slightly flatter density structures \citep{fvdt00}.

The calculation of the radio spectrum of gravitationally bound \hii\
regions is given by \citet{keto03}. Appendix~\ref{app:a} summarizes the
derivation, with a few errors in the formulae corrected.
The dotted lines in Figure~\ref{f:spec} are flux densities calculated
using these corrected formulae.

The only free parameter in the calculation is $n_0$, the density at
the molecular sonic point $r_m$. Adopting $T$=30~K as representative for the molecular
gas, the location of its sonic point only depends on the mass of the
central star. We have fitted the parameter $n_0$ to the observed (unweighted) flux
densities of our sources at frequencies up to 43~GHz.

For W~33A, we find $n_0$=3$\times$10$^5$ \ccm\ at $r_m$=0.49~pc,
for AFGL 2591, we find $n_0$=8$\times$10$^4$ \ccm\ at $r_m$=0.28~pc, and
for NGC 7538 IRS9, we find $n_0$=2$\times$10$^5$ \ccm\ at $r_m$=0.35~pc.
These densities agree to factors of a few with the values derived from
maps of \smm\ continuum and molecular lines \citep{fvdt00}.

Figure~\ref{f:spec} shows that the model reproduces the observed flux
densities to a factor of $\sim$2, which is the expected uncertainty
due to variability (in particular, variations in the accretion rate).
We conclude that ionized accretion flows describe our data well, but
that stronger tests are necessary to prove the model beyond doubt.

\subsection{Dust disks}
\label{s:disk}

As an alternative model, we consider the suggestion by
\citet{preibisch03} that the near-infrared emission from AFGL 2591 is
the inner rim of a thick disk or envelope at the dust sublimation
radius. These authors resolve the near-infrared emission and derive a
diameter of 40~AU, which is in good agreement with our measured 43~GHz
size, assuming a beam filling factor of unity.
Since our measured 43~GHz brightness is 850~K rather than 1500~K, this
interpretation requires either the optical depth to be $\approx$0.5 at
43~GHz, or the dust temperature to be $\approx$850~K, about half the
sublimation temperature.

The luminosity of a $T$=1500~K, 40~AU diameter black body sphere is
80,000~\lsol, which exceeds the observed value of 20,000~\lsol\
(\S~\ref{s:intro}). Therefore, if the object has this temperature and
size, it can fill at most 1/4 of the sky, and is likely a disk. However,
if $T$=850~K, the spherical black body luminosity is only 9000~\lsol,
within the observational limits.

The requirement of unit optical depth at 43~GHz constrains the dust
column density. Extrapolating Model~5 of the dust models by \citet{oh94},
which provides a good fit to the mid- to far-infrared spectrum of AFGL
2591 \citep{fvdt99}, the 43~GHz opacity is 0.06 cm$^2$\,g$^{-1}$. The
2.2~\mic\ opacity is a factor of 80,000 higher in this dust model, so that
$\tau$(43~GHz)=1 implies $A_V$=890,000 or $N_H$=1.8$\times$10$^{27}$
\scm. This column density is orders of magnitude higher than measured
for the envelopes of embedded high-mass stars (\citealt{fvdt00}; \citealt{hvdt03}),
but may be reached by massive circumstellar disks.

Another check of the disk model lies in the 2.2~\mic\ flux density of
2.8~Jy observed toward AFGL 2591 (e.g., \citealt{aitken88}). The
compact component in the speckle observations contributes $\approx$60\% of
the 2.2~\mic\ flux density of AFGL 2591 in a 12$''$ field (T.\ Preibisch, priv.\
comm.). If the emission is a black body at $T$=1500~K with a radius
$r$=20~AU, matching the observed brightness requires a foreground
2.2~\mic\ extinction of 25 magnitudes, or 20 mag if $T$=850~K. The
corresponding visual extinctions of 200--250 imply
$N_H$=2.0--2.5$\times$10$^{23}$ \scm, which agrees very well with the
$N$(H$_2$) = 2.3$\times$10$^{23}$ \scm\ measured in \smm\
emission \citep{fvdt99}. We conclude that the dust disk seen at
2.2~\mic\ by \citet{preibisch03} may also be responsible for our
observed 43~GHz emission.
The evidence for this interpretation would be strengthened by
measuring the orientation of the near-infrared emission, which the
current speckle data do not constrain (T.\ Preibisch, priv.\ comm.).

\subsection{Relation with large-scale outflows}

The deconvolutions in \S~\ref{s:possiz} indicate flattened shapes for
the radio sources, with axis ratios of $\approx$3.
One central prediction of disk accretion models of low-mass star formation
is that the outflow axis is perpendicular to the disk plane.
To see if such models may also work for high-mass star formation, this
section compares the orientation of the 43~GHz emission with that of
the large-scale outflows observed in our sources.

It is difficult to constrain the outflow orientation of such objects
from the commonly observed CO 1--0 line.
The outflows appear poorly collimated in single-dish observations, and
interferometric observations tend to resolve out most of the emission.
For example, in AFGL 2591, one of the earliest known outflow sources,
high velocity CO in the 1--0 line is spread over a large area
($\sim$5$'$ or $\sim$1~pc) and is poorly collimated \citep{lada84}.
Better constraints come from higher$-J$ CO lines
or other molecules such as HCO$^+$ \citep{beuther:outflow}.

In the case of AFGL 2591, CO 3--2 and HCO$^+$ 4--3 mapping by
\citet{hasegawa95} shows an outflow of size 90$\times$20$''$, embedded
in an arcminute-scale flow seen before in lower-$J$ CO lines. The
East-West orientation agrees with the positions of Herbig-Haro objects
and spots of shock-excited H$_2$ emission.
For NGC 7538 IRS9, maps of the CO 2--1 line \citep{mitchell:outflow}
show an outflow with a North-South orientation (PA$\sim$160$^\circ$).
Very high velocities (up to 110~\kms) are seen in the 3--2 line, but
this emission is spatially unresolved  \citep{mitchell:irs9}.
For W~33A, outflow activity is demonstrated by the positions and
velocities of OH and \hho\ masers (e.g., \citealt{fish03}) and by
`wings' on the \smm\ emission line profiles of CS, HCO$^+$ and other
molecules (van der Tak et al.\ 2000, 2003). Unfortunately, there is no
CO or HCO$^+$ map to give us the outflow orientation for this source.
In fact, the outflow may not be bipolar, as OH masers usually appear in
quasi-spherical outflows. Observations of \hho\ masers with 0.5~mas
resolution have revealed spherical outflows from several other young
high-mass stars (e.g., \citealt{torrelles:w75n}).

The 43~GHz emission in both AFGL 2591 and NGC 7538 IRS9 has a similar
orientation on the sky as their large-scale outflows. This alignment
argues against an origin of the 43~GHz emission in a disk,
%
%The of the 43~GHz emission with the large-scale outflow
and suggests an origin in a jet. However, thermal emission from an
ionized jet such as seen in lower-mass objects (e.g.,
\citealt{trinidad:lkha234}) is inconsistent with the measured spectral
indices (\S~\ref{s:spectra}). Perhaps cool material entrained by the
jet contains sufficient dust to produce the observed spectrum.

%\subsection{Jonathan Tan's jet model?}
\subsection{Photo-evaporating disks}

Another potential source of our observed 43~GHz emission is free-free emission
from an ionized wind from a circumstellar disk that is being illuminated by
the Lyman continuum of a hot central star.
Such photo-evaporating disks have been described semi-analytically by
\citet{hollenbach94} while numerical calculations were performed by
\citet{richling97}. 
In these models, evaporation occurs in the outer disk, where the radiative
force of the star is stronger than its gravitational force.
In our case of a $\sim$20~\msol\ star, this radius occurs at $\sim$100~AU,
which is comparable to our measured source size.

The models distinguish `strong' and `weak' stellar winds, depending on
whether the ram pressure of the wind is comparable to the thermal
pressure of the disk atmosphere or exceeds it. Wind speeds up to
several 100 \kms\ have been measured in mid-infrared CO absorption and
H emission lines toward AFGL 2591 and NGC 7538 IRS9
(\citealt{mitchell:outflow}; \citealt{bunn95}) and may also apply to
W~33A. For a wind speed of 500~\kms, a stellar mass of 20~\msol\ and a
Lyman continuum flux of 3$\times$10$^{48}$~s$^{-1}$
(\S~\ref{s:stevol}), the critical mass loss rate is 2$\times$10$^{-7}$
\msol\,yr$^{-1}$. The mass loss rates estimated from CO 2--1 are
$\sim$10$^{-4}$ \msol\,yr$^{-1}$, so the sources are clearly in the
strong wind limit.

The predicted spectral shape of circumstellar disks that are being
photo-evaporated by strong stellar winds is that of an optically thin \hii\
region ($\gamma$=$-$0.1) at short wavelengths, a thick \hii\ region ($\gamma$=2)
at long wavelengths, and a `wind' spectrum ($\gamma$=0.6) at intermediate
wavelengths. For the high mass loss rates of our sources, the transition to
optically thick emission lies at unobservably long wavelengths. The transition
from $\gamma$=0.6 to $\gamma$=$-$0.1 is expected around 0.5~cm
wavelength. However, the observed spectra have quite different shapes, so that
the photo-evaporating disk model does not seem applicable to our sources.

\subsection{Free electron absorption}
\label{s:mira}

As a last option, we consider free-free absorption by H$^-$ as source
of the observed 43~GHz emission. This mechanism was originally
proposed for the radio emission of Mira-type variable stars
\citep{reid:mira} and may play a role in Orion-KL \citep{beuther:orion}.
The required free electrons come from photo-ionization of Na and K
atoms, so that this mechanism only operates inside the dust
sublimation radius.

Our observed brightness temperatures of $\sim$1500~K are consistent
with this model, and combining the radii of $\sim$10~AU with
$N$(\hh)$\sim$10$^{27}$~\scm\ gives a density of $\sim$10$^{13}$~\ccm,
high enough to provide unit free electron opacity \citep{reid:mira}.
The broad-band radio spectra of W~33A and AFGL 2591 are not
inconsistent with the predicted $\gamma$=1.86 behaviour if the flux
densities are uncertain by a factor of $\sim$2. For NGC 7538 IRS9,
free electron absorption may explain the 15 -- 107~GHz spectrum, while
the lower-frequency data must have a different origin.
In any case, stronger tests of this model will come from measurements
of the density on $\sim$10~AU scales.

\section{Conclusions}
\label{conc}

% favour accretion over coagulation
We have detected weak, compact radio emission from three high-mass
protostars which are deeply embedded in molecular envelopes. In two
cases, the
emission is probably due to bremsstrahlung from gravitationally bound
\hii\ regions. The derived densities are in good agreement with values
found for the large-scale molecular envelope. Similar results were
found for another object, W3~IRS5 \citep{keck05}, and
additional support for accretion onto high-mass protostars comes from
\hho\ maser mapping of AFGL 2136 \citep{mvdt04}.
These findings support a picture of high-mass star formation where the
central star builds up mass by accretion from a molecular envelope. At
small radii, the stellar Lyman continuum ionizes the accretion flow,
but does not stop it.  As the star gains mass, its surface temperature
and ionizing flux increase.  The circumstellar ionized region grows
quasistatically until its radius reaches the critical value of
$GM/2c_s^2$. Stellar gravity is then unable to confine it, so that the
\hii\ region rapidly expands and halts the accretion.

However, this interpretation is not unique and may not be complete.
For AFGL 2591, the data are also consistent with dust emission from
a disk seen previously in the near-infrared.
Furthermore, NGC 7538 IRS9 appears to have a non-thermal cm-wave
spectrum, such as produced by synchrotron emission from a jet.
The mm-wave spectrum of this source is steeper than that of the
other two, suggesting optically thin dust where the others are thick.
In the case of W~33A, three radio continuum sources are seen at 3000 -- 5000
AU separations. This `clustering' of radio sources resembles the W3~IRS5
region ($d$=2.0~kpc, $L$=1.4$\times$10$^5$~\lsol), where seven sources
are found at about half the above separations \citep{keck05}.
Follow-up VLA observations would be useful to find out if some of the 43~GHz
sources in W~33A are transient, and mid-infrared imaging would reveal which of the
three are self-luminous. 
Together with our previous `case studies' of AFGL 2136
\citep{mvdt04} and W3~IRS5 \citep{keck05}, the current
observations suggest that the formation of stars and stellar
clusters with luminosities up to $\sim$10$^5$~\lsol\ 
proceeds through accretion and produces collimated outflows
as in the case of solar-type stars.  Such regions may well
contain small star clusters \citep{testi99} and clusters of
non-thermal radio sources \citep{gomez:ggd14}.  The main
difference in the formation of $\sim$10$^4$~\lsol\ stars
with that of solar-type stars is that their accretion flows
become ionized at small radii. In contrast, regions of
$\sim$10$^5$~\lsol\ appear to produce clusters of \hii\ 
regions and essentially uncollimated outflows, possibly as
the result of mergers of protostars or pre-stellar cores.

% future tests of model
In the future, the ionized accretion flow model should be tested by
measuring the density profiles of hypercompact \hii\ regions directly
through radio continuum observations at $\sim$10~AU resolution (10 mas
at 1~kpc) and $\sim$0.1~mJy sensitivity.
The duration of the ionized accretion flow phase should be measured by
a radio continuum survey of high-mass protostars at a resolution and
sensitivity comparable to this work.
Observations of radio recombination lines at high resolution exist for
one source \citep{keto:g10.6} and should be carried out for more.
These efforts probably exceed the capabilities of the VLA, but will be
feasible with the e-VLA (after 2006) and ALMA (after 2008).

\begin{acknowledgement}
  The authors thank Friedrich Wyrowski, Mark Reid and Thomas Preibisch for
  useful discussions, and the staff of the VLA (especially Claire Chandler)
  for assisting with the observations.
\end{acknowledgement}

\bibliographystyle{aa}
\bibliography{vla}

\appendix

\section{Flux density of an ionized accretion flow}
\label{app:a}

An H~II region with a density gradient consists of an optically thick
core and an optically thin envelope.
The angular radius of the optically thick core $\theta_1$ is related
to the angular radius of the \hii\ region $\theta_0$ by

$$
\theta_1^2 = 2 B^2 \left( -1 + \sqrt{1 + \theta_0^2 / B^2} \right)
$$
with
$$
B = A_{FF} n_0^2 T_e \theta_0^2 \frac{d}{{\rm pc}}
$$
and
$$
A_{FF} = 8.235 \times 10^{-2} T_e^{-0.35}
         \left( \frac{\nu}{{\rm GHz}} \right)^{-2.1}
$$

The radio emission from a gravitationally bound \hii\ region is the
sum of the emissions from the optically thick core and the optically
thin envelope, weighted by their respective areas on the sky.
For the optically thick core, the brightness temperature is
$$T_{\rm thick} = T_e \frac{\theta_1^2}{\theta_0^2}$$
while for the optically thin envelope,
$$
T_{\rm thin} = -4 A_{FF} n_0^2 \frac{d}{{\rm pc}}
\left( u + \theta_0/2 \ln \frac{\theta_0 - u}{ \theta_0 + u} \right)
$$
with
$$
u = \sqrt{\theta_0^2 - \theta_1^2}
$$

If the emission is unresolved, the total flux density is
$$S_\nu = 2 k_B (T_{\rm thick} + T_{\rm thin}) \Omega \nu^2 / c^2 $$
where $$\Omega = \pi \theta_0^2$$ is the solid angle of the \hii\ region.

\end{document}

%% file: defs.tex
\def\hii{H~{\sc II}}

\def\hh{H$_2$}

\def\hho{H$_2$O}

\def\smm{sub-mil\-li\-me\-ter}
\def\sas{sub-arc\-se\-cond}

\def\farcs{\mbox{$.\!\!^{\prime\prime}$}}

\def\gtsim{{_>\atop{^\sim}}}
\def\ltsim{{_<\atop{^\sim}}}

\def\kms{km~s$^{-1}$}

\def\scm{cm$^{-2}$}
\def\ccm{cm$^{-3}$}

\def\mic{$\mu$m}

\def\msol{M$_{\odot}$}
\def\lsol{L$_{\odot}$}

\def\teff{$T_{\rm eff}$}

\def\aap{{A\&A}}

\def\apjl{{ApJ}}
\def\apjs{{ApJS}}

\renewcommand{\citep}[1]{(\citeauthor{#1} \citeyear{#1})}

%% file: aa2872.bbl
\begin{thebibliography}{59}
\expandafter\ifx\csname natexlab\endcsname\relax\def\natexlab#1{#1}\fi

\bibitem[{{Aitken} {et~al.}(1988){Aitken}, {Smith}, {James}, {Roche}, \&
  {Hough}}]{aitken88}
{Aitken}, D.~K., {Smith}, C.~H., {James}, S.~D., {Roche}, P.~F., \& {Hough},
  J.~H. 1988, \mnras, 230, 629

\bibitem[{{Beuther} {et~al.}(2002){Beuther}, {Schilke}, {Sridharan}, {Menten},
  {Walmsley}, \& {Wyrowski}}]{beuther:outflow}
{Beuther}, H., {Schilke}, P., {Sridharan}, T.~K., {et~al.} 2002, \aap, 383, 892

\bibitem[{{Beuther} {et~al.}(2004){Beuther}, {Zhang}, {Greenhill}, {Reid},
  {Wilner}, {Keto}, {Marrone}, {Ho}, {Moran}, {Rao}, {Shinnaga}, \&
  {Liu}}]{beuther:orion}
{Beuther}, H., {Zhang}, Q., {Greenhill}, L.~J., {et~al.} 2004, \apjl, 616, L23

\bibitem[{{Bonnell} {et~al.}(1998){Bonnell}, {Bate}, \& {Zinnecker}}]{bbz98}
{Bonnell}, I.~A., {Bate}, M.~R., \& {Zinnecker}, H. 1998, \mnras, 298, 93

\bibitem[{{Bunn} {et~al.}(1995){Bunn}, {Hoare}, \& {Drew}}]{bunn95}
{Bunn}, J.~C., {Hoare}, M.~G., \& {Drew}, J.~E. 1995, \mnras, 272, 346

\bibitem[{{Campbell}(1984)}]{campbell84}
{Campbell}, B. 1984, \apj, 287, 334

\bibitem[{{Churchwell}(2002{\natexlab{a}})}]{churchwell02:boulder}
{Churchwell}, E. 2002{\natexlab{a}}, in Hot Star Workshop III: The Earliest
  Stages of Massive Star Birth. Edited by Paul A. Crowther. (ASP), 3

\bibitem[{{Churchwell}(2002{\natexlab{b}})}]{churchwell02:araa}
{Churchwell}, E. 2002{\natexlab{b}}, \araa, 40, 27

\bibitem[{{Claussen} {et~al.}(1994){Claussen}, {Gaume}, {Johnston}, \&
  {Wilson}}]{claussen94}
{Claussen}, M.~J., {Gaume}, R.~A., {Johnston}, K.~J., \& {Wilson}, T.~L. 1994,
  \apjl, 424, L41

\bibitem[{{Crampton} {et~al.}(1978){Crampton}, {Georgelin}, \&
  {Georgelin}}]{crampton78}
{Crampton}, D., {Georgelin}, Y.~M., \& {Georgelin}, Y.~P. 1978, \aap, 66, 1

\bibitem[{{Fish} {et~al.}(2003){Fish}, {Reid}, {Argon}, \& {Menten}}]{fish03}
{Fish}, V.~L., {Reid}, M.~J., {Argon}, A.~L., \& {Menten}, K.~M. 2003, \apj,
  596, 328

\bibitem[{{Forster} \& {Caswell}(1999)}]{forster99}
{Forster}, J.~R. \& {Caswell}, J.~L. 1999, \aaps, 137, 43

\bibitem[{{G{\' o}mez} {et~al.}(2002){G{\' o}mez}, {Rodr{\'{\i}}guez}, \&
  {Garay}}]{gomez:ggd14}
{G{\' o}mez}, Y., {Rodr{\'{\i}}guez}, L.~F., \& {Garay}, G. 2002, \apj, 571,
  901

\bibitem[{{Garay} \& {Lizano}(1999)}]{gl99}
{Garay}, G. \& {Lizano}, S. 1999, \pasp, 111, 1049

\bibitem[{{Gibb} {et~al.}(2000){Gibb}, {Whittet}, {Schutte}, {Boogert},
  {Chiar}, {Ehrenfreund}, {Gerakines}, {Keane}, {Tielens}, {van Dishoeck}, \&
  {Kerkhof}}]{gibb00}
{Gibb}, E.~L., {Whittet}, D.~C.~B., {Schutte}, W.~A., {et~al.} 2000, \apj, 536,
  347

\bibitem[{{Hasegawa} \& {Mitchell}(1995)}]{hasegawa95}
{Hasegawa}, T.~I. \& {Mitchell}, G.~F. 1995, \apj, 451, 225

\bibitem[{{Hatchell} \& {van der Tak}(2003)}]{hvdt03}
{Hatchell}, J. \& {van der Tak}, F.~F.~S. 2003, \aap, 409, 589

\bibitem[{{Hollenbach} {et~al.}(1994){Hollenbach}, {Johnstone}, {Lizano}, \&
  {Shu}}]{hollenbach94}
{Hollenbach}, D., {Johnstone}, D., {Lizano}, S., \& {Shu}, F. 1994, \apj, 428,
  654

\bibitem[{{Kameya} {et~al.}(1990){Kameya}, {Morita}, {Kawabe}, \&
  {Ishiguro}}]{kameya90}
{Kameya}, O., {Morita}, K., {Kawabe}, R., \& {Ishiguro}, M. 1990, \apj, 355,
  562

\bibitem[{{Keto}(2002{\natexlab{a}})}]{keto:g10.6}
{Keto}, E. 2002{\natexlab{a}}, \apj, 568, 754

\bibitem[{{Keto}(2002{\natexlab{b}})}]{keto02}
---. 2002{\natexlab{b}}, \apj, 580, 980

\bibitem[{{Keto}(2003)}]{keto03}
---. 2003, \apj, 599, 1196

\bibitem[{{Kr\"ugel}(2003)}]{kruegel}
{Kr\"ugel}, E. 2003, {The physics of interstellar dust} (Bristol, UK: The
  Institute of Physics)

\bibitem[{{Lada} {et~al.}(1984){Lada}, {Thronson}, {Smith}, {Schwartz}, \&
  {Glaccum}}]{lada84}
{Lada}, C.~J., {Thronson}, H.~A., {Smith}, H.~A., {Schwartz}, P.~R., \&
  {Glaccum}, W. 1984, \apj, 286, 302

\bibitem[{{Maeder} \& {Meynet}(1989)}]{maeder89}
{Maeder}, A. \& {Meynet}, G. 1989, \aap, 210, 155

\bibitem[{{Menten} \& {Reid}(1995)}]{menten95}
{Menten}, K.~M. \& {Reid}, M.~J. 1995, \apjl, 445, L157

\bibitem[{{Menten} \& {van der Tak}(2004)}]{mvdt04}
{Menten}, K.~M. \& {van der Tak}, F.~F.~S. 2004, \aap, 414, 289

\bibitem[{{Mitchell} {et~al.}(1989){Mitchell}, {Curry}, {Maillard}, \&
  {Allen}}]{mitchell:gl2591}
{Mitchell}, G.~F., {Curry}, C., {Maillard}, J., \& {Allen}, M. 1989, \apj, 341,
  1020

\bibitem[{{Mitchell} \& {Hasegawa}(1991)}]{mitchell:irs9}
{Mitchell}, G.~F. \& {Hasegawa}, T.~I. 1991, \apjl, 371, L33

\bibitem[{{Mitchell} {et~al.}(1990){Mitchell}, {Maillard}, {Allen}, {Beer}, \&
  {Belcourt}}]{mitchell:hot+cold}
{Mitchell}, G.~F., {Maillard}, J., {Allen}, M., {Beer}, R., \& {Belcourt}, K.
  1990, \apj, 363, 554

\bibitem[{{Mitchell} {et~al.}(1991){Mitchell}, {Maillard}, \&
  {Hasegawa}}]{mitchell:outflow}
{Mitchell}, G.~F., {Maillard}, J.-P., \& {Hasegawa}, T.~I. 1991, \apj, 371, 342

\bibitem[{{Neufeld} \& {Melnick}(1991)}]{neufeld91}
{Neufeld}, D.~A. \& {Melnick}, G.~J. 1991, \apj, 368, 215

\bibitem[{{Olnon}(1975)}]{olnon75}
{Olnon}, F.~M. 1975, \aap, 39, 217

\bibitem[{{Ossenkopf} \& {Henning}(1994)}]{oh94}
{Ossenkopf}, V. \& {Henning}, T. 1994, \aap, 291, 943

\bibitem[{{Osterbrock}(1989)}]{osterbrock}
{Osterbrock}, D.~E. 1989, {Astrophysics of gaseous nebulae and active galactic
  nuclei} (Mill Valley, CA, University Science Books)

\bibitem[{{Ott} {et~al.}(1994){Ott}, {Witzel}, {Quirrenbach}, {Krichbaum},
  {Standke}, {Schalinski}, \& {Hummel}}]{ott94}
{Ott}, M., {Witzel}, A., {Quirrenbach}, A., {et~al.} 1994, \aap, 284, 331

\bibitem[{{Preibisch} {et~al.}(2003){Preibisch}, {Balega}, {Schertl}, \&
  {Weigelt}}]{preibisch03}
{Preibisch}, T., {Balega}, Y.~Y., {Schertl}, D., \& {Weigelt}, G. 2003, \aap,
  412, 735

\bibitem[{{Reid} {et~al.}(1995){Reid}, {Argon}, {Masson}, {Menten}, \&
  {Moran}}]{reid95}
{Reid}, M.~J., {Argon}, A.~L., {Masson}, C.~R., {Menten}, K.~M., \& {Moran},
  J.~M. 1995, \apj, 443, 238

\bibitem[{{Reid} \& {Menten}(1997)}]{reid:mira}
{Reid}, M.~J. \& {Menten}, K.~M. 1997, \apj, 476, 327

\bibitem[{{Rengarajan} \& {Ho}(1996)}]{rengarajan96}
{Rengarajan}, T.~N. \& {Ho}, P.~T.~P. 1996, \apj, 465, 363

\bibitem[{{Richling} \& {Yorke}(1997)}]{richling97}
{Richling}, S. \& {Yorke}, H.~W. 1997, \aap, 327, 317

\bibitem[{{Sandell} {et~al.}(2005){Sandell}, {Goss}, \& {Wright}}]{sandell05}
{Sandell}, G., {Goss}, W.~M., \& {Wright}, M. 2005, \apj, 621, 839

\bibitem[{{Schaerer} \& {de Koter}(1997)}]{sdk97}
{Schaerer}, D. \& {de Koter}, A. 1997, \aap, 322, 598

\bibitem[{{Schutte} {et~al.}(1996){Schutte}, {Tielens}, {Whittet}, {Boogert},
  {Ehrenfreund}, {de Graauw}, {Prusti}, {van Dishoeck}, \&
  {Wesselius}}]{schutte96}
{Schutte}, W.~A., {Tielens}, A.~G.~G.~M., {Whittet}, D.~C.~B., {et~al.} 1996,
  \aap, 315, L333

\bibitem[{{Tan}(2003)}]{tan03}
{Tan}, J.~C. 2003, in ASP Conf. Ser. 287: Galactic Star Formation Across the
  Stellar Mass Spectrum, 207--218

\bibitem[{{Testi} {et~al.}(1999){Testi}, {Palla}, \& {Natta}}]{testi99}
{Testi}, L., {Palla}, F., \& {Natta}, A. 1999, \aap, 342, 515

\bibitem[{{Tieftrunk} {et~al.}(1997){Tieftrunk}, {Gaume}, {Claussen}, {Wilson},
  \& {Johnston}}]{tieftrunk97}
{Tieftrunk}, A.~R., {Gaume}, R.~A., {Claussen}, M.~J., {Wilson}, T.~L., \&
  {Johnston}, K.~J. 1997, \aap, 318, 931

\bibitem[{{Tofani} {et~al.}(1995){Tofani}, {Felli}, {Taylor}, \&
  {Hunter}}]{tofani95}
{Tofani}, G., {Felli}, M., {Taylor}, G.~B., \& {Hunter}, T.~R. 1995, \aaps,
  112, 299

\bibitem[{{Torrelles} {et~al.}(2003){Torrelles}, {Patel}, {Anglada}, {G{\'
  o}mez}, {Ho}, {Lara}, {Alberdi}, {Cant{\' o}}, {Curiel}, {Garay}, \&
  {Rodr{\'{\i}}guez}}]{torrelles:w75n}
{Torrelles}, J.~M., {Patel}, N.~A., {Anglada}, G., {et~al.} 2003, \apjl, 598,
  L115

\bibitem[{{Trinidad} {et~al.}(2003){Trinidad}, {Curiel}, {Cant{\' o}},
  {D'Alessio}, {Rodr{\'{\i}}guez}, {Torrelles}, {G{\' o}mez}, {Patel}, \&
  {Ho}}]{trinidad:gl2591}
{Trinidad}, M.~A., {Curiel}, S., {Cant{\' o}}, J., {et~al.} 2003, \apj, 589,
  386

\bibitem[{{Trinidad} {et~al.}(2004){Trinidad}, {Curiel}, {Torrelles},
  {Rodr{\'{\i}}guez}, {Cant{\' o}}, {G{\' o}mez}, {Patel}, \&
  {Ho}}]{trinidad:lkha234}
{Trinidad}, M.~A., {Curiel}, S., {Torrelles}, J.~M., {et~al.} 2004, \apj, 613,
  416

\bibitem[{{Van der Tak}(2003)}]{fvdt03}
{Van der Tak}, F.~F.~S. 2003, in IAU Symposium 221: Star Formation at High
  Angular Resolution, 59--66

\bibitem[{{Van der Tak} {et~al.}(2003){Van der Tak}, {Boonman}, {Braakman}, \&
  {van Dishoeck}}]{sulph03}
{Van der Tak}, F.~F.~S., {Boonman}, A.~M.~S., {Braakman}, R., \& {van
  Dishoeck}, E.~F. 2003, \aap, 412, 133

\bibitem[{{Van der Tak} {et~al.}(2005){Van der Tak}, {Tuthill}, \&
  {Danchi}}]{keck05}
{Van der Tak}, F.~F.~S., {Tuthill}, P.~G., \& {Danchi}, W. 2005, \aap, 431, 993

\bibitem[{{Van der Tak} {et~al.}(1999){Van der Tak}, {van Dishoeck}, {Evans},
  {Bakker}, \& {Blake}}]{fvdt99}
{Van der Tak}, F.~F.~S., {van Dishoeck}, E.~F., {Evans}, N.~J., {Bakker},
  E.~J., \& {Blake}, G.~A. 1999, \apj, 522, 991

\bibitem[{{Van der Tak} {et~al.}(2000){Van der Tak}, {van Dishoeck}, {Evans},
  \& {Blake}}]{fvdt00}
{Van der Tak}, F.~F.~S., {van Dishoeck}, E.~F., {Evans}, N.~J., \& {Blake},
  G.~A. 2000, \apj, 537, 283

\bibitem[{{Van Dishoeck}(1998)}]{ewine98}
{Van Dishoeck}, E.~F. 1998, in Chemistry and Physics of Molecules and Grains in
  Space. Faraday Discussions No. 109, 31

\bibitem[{{Walmsley}(1995)}]{walmsley95}
{Walmsley}, C.~M. 1995, Rev. Mexicana Astron. Astrofis. Ser. de Conf., 1, 137

\bibitem[{{Willner} {et~al.}(1982){Willner}, {Gillett}, {Herter}, {Jones},
  {Krassner}, {Merrill}, {Pipher}, {Puetter}, {Rudy}, {Russell}, \&
  {Soifer}}]{willner82}
{Willner}, S.~P., {Gillett}, F.~C., {Herter}, T.~L., {et~al.} 1982, \apj, 253,
  174

\end{thebibliography}
